\documentclass[aps,pra,twocolumn,floats]{revtex4-2} 
\usepackage{amsmath}
\usepackage{graphicx}
\usepackage{dcolumn}
\usepackage{bm}
\usepackage{amssymb}
\usepackage{latexsym}
\usepackage{color}
\usepackage{subfigure}
\usepackage{ulem}

\begin{document}

\title{The Aharonov-Casher Phase: Considerations Regarding Force, Time-Dependence, and Berry Phase}

\author{Igor Kuzmenko$^{1,2}$, Y. B. Band$^{1,2,3}$, Yshai Avishai$^{1,4}$}

\affiliation{
  $^1$Department of Physics,
  Ben-Gurion University of the Negev,
  Beer-Sheva 84105, Israel
  \\
  $^2$Department of Chemistry,
  Ben-Gurion University of the Negev,
  Beer-Sheva 84105, Israel
  \\
  $^3$The Ilse Katz Center for Nano-Science,
  Ben-Gurion University of the Negev,
  Beer-Sheva 84105, Israel
  \\
  $^4$Yukawa Institute for Theoretical Physics, Kyoto, Japan
  }

\begin{abstract}
The relation of the Aharonov-Casher (AC) effect and the force on a particle having a magnetic moment is explored.  The general form of the AC Hamiltonian is derived using the Foldy-Wouthuysen transformation to the Dirac equation.  Geometries in which an analytic expression for the phase can be obtained are examined, as well as the relation of the AC phase to the Berry phase.  The AC phase is determined for an arbitrary homogeneous electric field; it is quadratic (linear) in the field strength for small (large) electric field strengths.
\end{abstract}

\maketitle

\section{Introduction}   \label{sec:Intro}

The Aharonov-Casher (AC) effect \cite{AC_84}, originally proposed by Aharonov and Casher in 1984, is a fundamental quantum phenomenon that elucidates the role of geometric phases in quantum mechanics.  It demonstrates how the presence of an electric field can affect the quantum states of particles having a magnetic moment.  Specifically, the AC effect demonstrates that the wave function of a particle with a magnetic moment moving in an external electric field acquires a geometric phase that is, in general, dependent on the path taken and the properties of the field.  This leads to observable consequences in interference patterns.  The implications of the AC effect extend beyond theoretical interest, finding applications in various domains of quantum mechanics, particularly in the field of interferometry. The AC effect can be harnessed in particle interferometers to manipulate and control the phase of particle waves, thus providing a mechanism for precise measurements of physical quantities, such as the magnetic dipole moment of atoms and particles \cite{Cimmino_89}. Advancements in experimental techniques have enabled the observation of the AC effect with greater precision \cite{Gillot_14}.  According to the aforementioned reference, at least six experiments measured the AC effect prior to 2015. 

In a recent Letter~\cite{ACP}, we demonstrated that the AC phase is a geometric phase that, in general, depends on the details of the closed path taken by a particle with a magnetic moment that is subject to an electric field.  Consequently, it is not a topological phase. The proof of this statement is obtained by developing a counterexample that elucidates the dependence of the AC phase on the details of the path. Furthermore, Ref.~\cite{ACP} demonstrated that paths having an Abelian AC phase factor, can also have an AC phase that is path-independent, whereas paths having a non-Abelian AC phase factor may have an AC phase that is path-dependent (i.e., not topological).

This paper aims to clarify the following issues regarding the AC phase: (1) the relation of the AC phase to classical force on the particle, (2) the general form of the AC Hamiltonian, (3) the AC phase in a homogeneous electric field, (4) the relation of the AC phase and the Berry phase. Section~\ref{sec:Center} calculates the force exerted on a neutral particle that possesses a magnetic moment, and that moves in a planar ring in the presence of an electric field generated by a line of charge with linear charge density $\lambda$, that passes through the center of the ring and tilted at an angle $\theta$ from the normal to the plane of the ring.  Section~\ref{sec:General_AC_Hamiltonian} derives the Hamiltonian for the Aharonov-Casher effect \cite{AC_84} by applying the Foldy-Wouthuysen transformation to the Dirac equation \cite{Frohlich_93, BD, Band-Avishai-QuantumMechanics}, and shows that, in general, the Hamiltonians used in both Refs.~\cite{Peshkin_95} and \cite{Dulat_12} need to be amended.  Section \ref{sec:not-center} considers the Aharonov-Casher phase $\varphi_{\rm AC}$ for a ring threaded by the line of charge oriented perpendicular to the plane of the ring, and intersecting the plane at a distance $R$ from the center of the ring.  In Sec.~\ref{sec:AC_phase_Berry_phase} we demonstrate that the AC phase is a special case of a Berry phase, and Sec.~\ref{sec:homogeneous-E-field} analyses the AC phase of a particle with spin-1/2 along a circular path and placed in a homogeneous electric field.  It is shown that, generally, the Aharonov-Casher (AC) phase factor is non-Abelian and the AC phase is not topological.  Finally, Sec.~\ref{sec:Summary} contains a summary and conclusion.

\section{Force when Line of Charge passes through the center of the ring}   \label{sec:Center}

In cylindrical coordinates ${\bf r}=(r, \phi, z)$, the position of the particle propagating on a ring in the $x$-$y$ plane, having radius $r_0$ is ${\bf r}=(r_0, \phi, 0) \equiv {\bf r}(\phi)$; in Cartesian coordinates,
\begin{equation}   \label{eq:r-phi}
  {\bf r}(\phi) = r_0 (\cos \phi \, \hat{\bf x} + \sin \phi \, \hat{\bf y}) ,
\end{equation}
where $0 \leq \phi < 2 \pi$.
The cylindrical basis vectors are ${\bf e}_{r}$, ${\bf e}_{\phi}$ and ${\bf e}_{z}$.  Applying the Foldy-Wouthuysen scheme to the Dirac equation \cite{Frohlich_93}, the following Hamiltonian for a neutral particle with spin 1/2 is obtained:
\begin{eqnarray}
H &=&
\frac{1}{2 M} \, {\bf p}^2 +
\frac{g \mu_B}{8 M c}
\big(
  {\bf p} \cdot [\boldsymbol\sigma \times {\bf E} ({\bf r})] +
  [\boldsymbol\sigma \times {\bf E} ({\bf r})] \cdot {\bf p}
\big)
\nonumber \\ && +
\left(\frac{\alpha}{2} + \frac{g^2 \mu_B^2}{8 M c^2} \right) \, {\bf E}^{2} ({\bf r}) +
\frac{\hbar g \mu_B}{8 M c} \, {\boldsymbol \nabla} \cdot {\bf E} ({\bf r}) .
\label{eq:Foldy-Wouthuysen-expansion}
\end{eqnarray}
Here $M$ is the particle mass, $\boldsymbol\sigma = \sigma_x \hat{\bf x} + \sigma_y \hat{\bf y} + \sigma_z \hat{\bf z}$ is the Pauli spin vector, $\mu_B = \tfrac{e \hbar}{2 m_e c}$ is the Bohr magneton in Gaussian units, $g$ is the gyromagnetic ratio, $m_e$ is the electron mass, $e$ is the elementary charge, $c$ is the speed of light, and $\alpha$ is the particle polarizability.  For the particle moving on a ring of radius $r_0$ placed in the $x$-$y$ plane, the momentum operator is ${\bf p} =  {\bf e}_{\phi} p$, where $p = - \frac{i \hbar}{r_0} \partial_{\phi}$.   According to Gauss's law, ${\boldsymbol \nabla} \cdot {\bf E} ({\bf r},\theta) = 4 \pi \rho ({\bf r},\theta)$, where $\rho ({\bf r},\theta) = \lambda \delta(x \cos \theta - z \sin \theta) \delta(y)$ is the charge density of the line of charge, $\lambda$ is the linear charge density of the line of charge, $\delta (\bullet)$ is the Dirac delta function, and $\theta$ is the angle between the line of charge and the $z$-axis.  For ${\bf r} = {\bf r} (\phi)$, $[ {\boldsymbol \nabla} \cdot {\bf E} ({\bf r},\theta) ] = 0$, and the Hamiltonian in Eq.~(\ref{eq:Foldy-Wouthuysen-expansion}) can be expressed as
\begin{eqnarray}   \label{eq:H-res}
H &=&
\frac{1}{2 M} \,
\Big( p + \frac{g \mu_B}{4 c} \, \boldsymbol\sigma \cdot [ {\bf E} ({\bf r} (\phi),\theta) \! \times \! {\bf e}_{\phi} ] \Big)^{2} + \frac{\alpha}{2} {\bf E}^{2} ({\bf r} (\phi),\theta) 
\nonumber \\ && +
\frac{g^2 \mu_B^2}{8 M c^2} \,
\Big( {\bf E}^{2} ({\bf r} (\phi),\theta) - \frac{1}{4} \, [ {\bf E} ({\bf r} (\phi),\theta) \times {\bf e}_{\phi} ]^{2} \Big).
\end{eqnarray}
For the case of a particle propagating on a ring threaded by a line of charge that passes {\it through the center of the ring}, as discussed in Ref.~\cite{ACP}, the electric field is given by ${\bf E} ({\bf r} (\phi), \theta) = \tfrac{\lambda}{r_0} \boldsymbol{\mathcal E} (\phi, \theta)$, where the dimensionless electric field $\boldsymbol{\mathcal E} (\phi, \theta)$ is given by
\begin{eqnarray}
  \boldsymbol{\mathcal E} (\phi, \theta) &=&
  2 \, {\bf e}_{r} +
  \frac{\sin^2 \theta \, \sin (2 \phi)}{\cos^2 \theta \, \cos^2 \phi + \sin^2 \phi} \, {\bf e}_{\phi}
  \nonumber \\ && -
  \frac{\sin (2 \theta) \, \cos \phi}{\cos^2 \theta \, \cos^2 \phi + \sin^2 \phi} \, {\bf e}_{z} .
  \label{eq:E-line}
\end{eqnarray}
The Hamiltonian in Eq.~(\ref{eq:H-res}) becomes
\begin{eqnarray}   \label{eq:H_AC}
H (\theta) &=&
\frac{1}{2 M} \,
\Big( p + \frac{\hbar}{r_0} \, {\boldsymbol \sigma} \cdot \boldsymbol {\mathcal{A}} (\phi, \theta) \Big)^{2}
+ \frac{\alpha \, \lambda^2}{2 r_0^2} \boldsymbol{\mathcal E}^2 (\phi, \theta)
\nonumber \\ && +
\frac{\hbar^2}{8 M r_0^2} \,
\frac{\lambda^2}{\lambda_0^2} \,
\Big(
  3 \, \mathcal{E}^{2} (\phi, \theta) +
  \mathcal{E}_{\phi}^{2} (\phi, \theta)
\Big) ,
\end{eqnarray}
where $\mathcal{E}_{r} (\phi, \theta)$, $\mathcal{E}_{\phi} (\phi, \theta)$ and $\mathcal{E}_{z} (\phi, \theta)$ are cylindrical components of $\boldsymbol{\mathcal E} (\phi, \theta)$, $\mathcal{E} (\phi, \theta) = |\boldsymbol{\mathcal E} (\phi, \theta)|$, the dimensionless vector potential $\boldsymbol {\mathcal{A}} (\phi, \theta)$ is given by \cite{ACP}, 
\begin{eqnarray}
  {\boldsymbol {\mathcal{A}}} (\phi, \theta) &=&
  \frac{g \mu_B \lambda}{4 \hbar c} \, \boldsymbol{\mathcal E} (\phi, \theta)  \times{\bf e}_{\phi} =
  \frac{\lambda}{\lambda_0}
  \bigg[
    {\bf e}_{z}
    \nonumber \\ && +
    \frac{1}{2} \,
    \frac{\sin (2 \theta) \, \cos (\phi)}{\cos^2 (\theta) \, \cos^2 (\phi) + \sin^2 (\phi)} \,
    {\bf e}_{r}
  \bigg] .
  \label{eq:vector-potential-SU2}
\end{eqnarray}
and
\begin{equation}   \label{eq:lambda_0}
  \lambda_0 = \frac{2 \hbar c}{g \mu_B} .
\end{equation}
The force acting on the particle in the direction ${\bf e}_{\phi}$ is given by
\begin{equation}   \label{eq:force-def}
\mathcal{F} (\phi, \theta) =
\frac{d \, \Pi (\phi)}{d t} =
\frac{i}{\hbar} \, \big[ H (\theta), \Pi (\phi, \theta) \big] ,
\end{equation}
where $H (\theta)$ is given in Eq.~(\ref{eq:H_AC}), and the canonical momentum operator in the direction ${\bf e}_{\phi}$, $\Pi (\phi, \theta)$, is
\begin{eqnarray}   \label{eq:momentum-def}
\Pi (\phi, \theta) &=&
M r_0 \, \frac{d \phi}{d t} = \frac{i M r_0}{\hbar} \, \big[ H (\theta), \phi \big] \nonumber \\
 &=& p + \frac{\hbar}{r_0} \, {\boldsymbol \sigma} \cdot {\boldsymbol {\mathcal{A}}}(\phi, \theta) .
\end{eqnarray}
Calculation of the commutator $[H (\theta), \Pi (\phi, \theta)]$ in Eq.~(\ref{eq:force-def}) yields
\begin{eqnarray}   \label{eq:force-res}
\mathcal{F} (\phi, \theta) &=&
- \frac{\hbar^2}{8 M r_0^2} \,
\frac{\lambda^2}{\lambda_0^2} \,
\frac{d}{d \phi}
\big[
  4 \mathcal{E}_{\phi}^{2} (\phi, \theta) +
  3 \mathcal{E}_{z}^{2} (\phi, \theta)
\big]
\nonumber \\ && -
\frac{\alpha \, \lambda^2}{2 r_0^2} \,
\frac{d}{d \phi}
\big[
  \mathcal{E}_{\phi}^{2} (\phi, \theta) +
  \mathcal{E}_{z}^{2} (\phi, \theta)
\big] ,
\end{eqnarray}
since the component $\mathcal{E}_r(\theta) = 2$ is independent of $\phi$, therefore $d \mathcal{E}_r^2 / d \phi = 0$.   Furthermore, the component $\mathcal{E}_z (\phi, \theta)$ is proportional to $\sin (2 \theta)$, and the component $\mathcal{E}_{\phi}  (\phi, \theta)$ is proportional to $\sin^2 \theta$.  The force is therefore given by
\begin{eqnarray} \label{eq:force}
\mathcal{F} (\phi, \theta) &=&
- \frac{\hbar^2}{4 M r_0^2} \, \frac{\lambda^2}{\lambda_0^2} \,
\frac{\sin^2 \theta}{(\cos^2 \theta \, \cos^2 \phi + \sin^2 \phi)^3}
\nonumber \\ && \times
\big\{
  [39 + 8 \cos (2 \theta) + \cos (4 \theta)] \, \sin (2 \phi)
  \nonumber \\ && -
  2 \, [9 + \cos (2 \theta)] \, \sin^2 \theta \, \sin (4 \phi)
\big\}
\nonumber \\ && +
\frac{\alpha \, \lambda^2}{r_0^2} \,
\frac{2 \, \sin^2 \theta \, \sin (2 \phi)}{\big( \cos^2\theta \, \cos^2 \phi + \sin^2 \phi \big)^{2}} .
\end{eqnarray}

\begin{figure}
\centering
  \includegraphics[width=0.9\linewidth,angle=0] {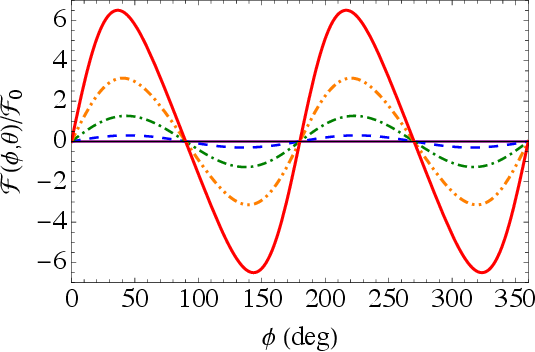}
\caption{\footnotesize Force $\mathcal{F} (\phi, \theta)$ versus the angle $\phi$ for a variety of values of the angle $\theta$: (a) $\theta = 0$, (b) $\theta = \pi / 20$, (c) $\theta = \pi / 10$, (d) $\theta = 3 \pi / 20$ and (e) $\theta = \pi / 5$.  The force is given in units of $\mathcal{F}_{0} = \frac{\hbar^2}{4 M r_0^2} \, \frac{\lambda^2}{\lambda_0^2}$.  The force is calculated for a neutron.  The polarizability of a neutron is small and its effect on the force is very small on the scale used in the figure.
}
\label{Fig:Force-AC}
\end{figure}

Figure~\ref{Fig:Force-AC} plots the force $\mathcal{F} (\phi, \theta)$ acting on a neutron versus the angle $\phi$ for a variety of values of the angle $\theta$.  The force vanishes if $\theta = 0$. For $\theta \neq 0$, the force is periodic function of $\phi$ with period $\pi$, and the integral of the force over the whole path vanishes, $\int\limits_{0}^{2 \pi} \mathcal{F} (\phi, \theta) \, d \phi = 0$.

\section{General AC Hamiltonian for Time-Dependent Fields}   \label{sec:General_AC_Hamiltonian}

The general Pauli Hamiltonian derived from the Dirac Hamiltonian for the spin-1/2 particle by making the Foldy-Wouthuysen transformation \cite{Frohlich_93, BD, Band-Avishai-QuantumMechanics} is
\begin{eqnarray}
H &=&
\frac{1}{2 M} \, {\bf p}^2 +
\frac{g \mu_B}{8 M c}
\big(
  {\bf p} \cdot [\boldsymbol\sigma \times {\bf E} ({\bf r}, t)] +
  [\boldsymbol\sigma \times {\bf E} ({\bf r}, t)] \cdot {\bf p}
\big) 
\nonumber \\ && +
\frac{\tilde \alpha}{2} \, {\bf E}^{2} ({\bf r}, t) +
\frac{\hbar g \mu_B}{8 M c} \, {\boldsymbol \nabla} \cdot {\bf E} ({\bf r}, t) - {\boldsymbol \mu} \cdot {\bf B}({\bf r}, t),
\label{eq:Foldy-Wouthuysen-Hamiltonian-general}
\end{eqnarray}
where ${\boldsymbol \mu} = \frac{1}{2} g \mu_B {\boldsymbol \sigma}$ is the magnetic moment operator for the particle and
\begin{equation} \label{eq:alpha-tilde}
\tilde \alpha = \alpha + \frac{g^2 \mu_B^2}{4 M c^2} .
\end{equation}

The Hamiltonian in Eq.~(7) in Ref.~\cite{Peshkin_95}, together with Eq.~(16) in Ref.~\cite{Peshkin_95} for the magnetic field experienced by the particle in its rest frame, yields the Peshkin-Lipkin Hamiltonian:
\begin{equation}   \label{eq:PL}
  H_{\rm PL} = \frac{{\bf p}^{2}}{2 M} -
  \frac{\mu}{M c} \,
  \boldsymbol\sigma \cdot \big[ {\bf p} \times {\bf E} ({\bf r}, t) \big] .
\end{equation}
Reference \cite{Dulat_12} employs the same Hamiltonian, but changed the order of the operators in the spin orbit term.  That is to say,  they use $\boldsymbol\sigma \cdot [{\bf E} ({\bf r}, t) \times  {\bf v}]$ in SI units. Reference~\cite{Dulat_12} asserts that Ref.~\cite{Peshkin_95} used the ``wrong Hamiltonian which yields their conclusion incorrect''. We shall now show that the Hamiltonian used by Ref.~\cite{Peshkin_95} and Ref.~\cite{Dulat_12} are equivalent; moreover, they are both the correct form for a time-dependent electric field.  The Hamiltonians in Refs.~\cite{Peshkin_95, Dulat_12} are Hermitian {\it only} if
\begin{eqnarray}   \label{eq:p-dot-s-cross-E=s-cross-E-dot-p}
  \boldsymbol\sigma \cdot \big[ {\bf p} \times {\bf E} ({\bf r}, t) \big] &\equiv&
  -{\bf p} \cdot [\boldsymbol\sigma \times {\bf E} ({\bf r}, t)] 
    \nonumber \\ &=&
  -[\boldsymbol\sigma \times {\bf E} ({\bf r}, t)] \cdot {\bf p} .
\end{eqnarray}
The first equality in Eq.~(\ref{eq:p-dot-s-cross-E=s-cross-E-dot-p}) is always true because $[\boldsymbol\sigma,{\bf p}] = 0$, but the second equality is generally false.  Indeed, the difference of the operators ${\bf p} \cdot [\boldsymbol\sigma \times {\bf E} ({\bf r}, t)]$ and $[\boldsymbol\sigma \times {\bf E} ({\bf r}, t)] \cdot {\bf p}$ is given by
\begin{eqnarray*}
  &&
  {\bf p} \cdot [\boldsymbol\sigma \times {\bf E} ({\bf r}, t)] -
  [\boldsymbol\sigma \times {\bf E} ({\bf r}, t)] \cdot {\bf p}
  =
  - i \hbar \, \nabla \cdot [\boldsymbol\sigma \times {\bf E} ({\bf r}, t)]
  \nonumber \\ && \qquad =
  - i \hbar \, \boldsymbol\sigma \cdot [\nabla \times {\bf E} ({\bf r}, t)] =
  \frac{i \hbar}{c} \, \boldsymbol\sigma \cdot \frac{\partial {\bf B} ({\bf r}, t)}{\partial t} ,
\end{eqnarray*}
where the final equality is due to Faraday's law (in Gaussian units).  If $\partial {\bf B} ({\bf r}, t)/\partial t = 0$, the equality (\ref{eq:p-dot-s-cross-E=s-cross-E-dot-p}) is correct, but both references \cite{Peshkin_95} and \cite{Dulat_12} consider time-dependent fields.  For a time-dependent electromagnetic field, $\partial {\bf B} ({\bf r}, t)/\partial t \neq 0$, therefore Eq.~(\ref{eq:Foldy-Wouthuysen-Hamiltonian-general}) should be used rather than Eq.~(\ref{eq:PL}).  Moreover, for a time-dependent charge density of the line of charge, $\lambda(t)$, the electric field is time-dependent, $\boldsymbol{\mathcal E} ({\bf r}, t)$, and the Hamiltonian in Eq.~(\ref{eq:Foldy-Wouthuysen-Hamiltonian-general}) is time-dependent.  This results in a dynamical AC effect which can be treated with the time-dependent Schr\"odinger equation or, if the rate of change of the charge density were slow, with the adiabatic approximation.  When the electromagnetic field is time-dependent, the AC effect will be accompanied with additional effects, such as Zeeman and Aharonov-Bohm effects.  We shall not pause to develop this here.

\section{Line of Charge off-center from the ring}   \label{sec:not-center}

\begin{figure}
\centering
  \includegraphics[width=0.8\linewidth,angle=0] {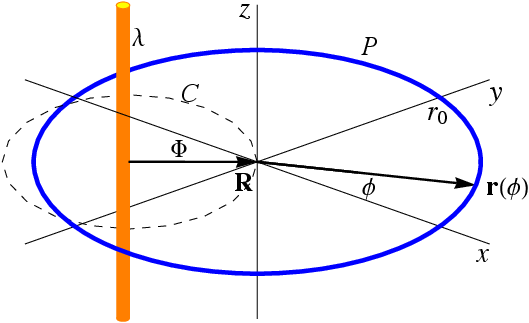}
\caption{\footnotesize Path $P$ encircles a line of charge with linear charge density $\lambda$.  The path is a circle of radius $r_0$, in the plane perpendicular to the line of charge and the center of the circle is at ${\bf R}$.  The position of a particle on the circle is ${\bf r} (\phi)$.}
\label{Fig:AC-phase-Berry-phase}
\end{figure}

We proceed by following Refs.~\cite{Reuter_91} and demonstrate that the AC phase is a special case of Berry's phase.  Let us consider the motion of a particle moving along a ring of radius $r_0$, placed in the $x$-$y$ plane and threaded by a line of charge with linear charge density $\lambda$.  The center of the ring is placed at ${\bf R} = R \, \cos \Phi \, \hat{\bf x} + R \, \sin \Phi \, \hat{\bf y}$, and the line of charge is parallel to the $z$-axis, as shown in Fig.~\ref{Fig:AC-phase-Berry-phase}.  The position of the particle on the ring relative to its center is ${\bf r}(\phi)$ defined in Eq.~(\ref{eq:r-phi}), where $r_0 > R$ (the line of charge threads the ring).  In this case, the electric field ${\bf E} ({\bf r} (\phi), {\bf R})$ generated by the line of charge on the ring is given by ${\bf E} ({\bf r} (\phi), {\bf R}) = \frac{\lambda}{r_0} \boldsymbol{\mathcal E} (\phi,{\bf  R})$, where the dimensionless electric field $\boldsymbol{\mathcal E} (\phi,{\bf  R})$ is given by
\begin{eqnarray}   \label{eq:EF-R}
\boldsymbol{\mathcal E} (\phi,{\bf  R}) &=&
\frac{2 r_0}{| {\bf r} (\phi) + {\bf R} |^2} \, \big( {\bf r} (\phi) + {\bf R} \big)
\nonumber \\ &=&
\frac{2 r_0 \, (r_0 + R \cos (\phi - \Phi)}{r_0^2 + R^2 + 2 r_0 R \cos (\phi - \Phi)} \, {\bf e}_{r}
\nonumber \\ &-&
\frac{2 r_0 \, R \sin (\phi - \Phi)}{r_0^2 + R^2 + 2 r_0 R \cos (\phi - \Phi)} \, {\bf e}_{\phi} .
\end{eqnarray}

The Hamiltonian $H({\bf R})$ of the particle is similar to $H$ in Eq.~(\ref{eq:H_AC}) but now the Hamiltonian depends on ${\bf R}$.  In this section, the term proportional to $\lambda^2/\lambda_0^2$ in Eq.~(\ref{eq:H_AC}) is neglected because $\frac{\lambda}{\lambda_0} = \frac{g \mu_B \lambda}{2 \hbar c} = \frac{g e \lambda}{4 m_e c^2} \ll 1$, therefore Eq.~(\ref{eq:H_AC}) is approximated by (see Ref.~\cite{Mignani_91})
\begin{eqnarray}   \label{eq:H-vs-R}
H ({\bf R}) &=&
\frac{1}{2 M} \, \Pi^{2} (\phi - \Phi, R) ,
\end{eqnarray}
where
\begin{eqnarray}
&&
\Pi (\phi - \Phi, R) =
p + \frac{\hbar}{r_0} \, {\boldsymbol \sigma} \cdot {\boldsymbol {\mathcal{A}}} (\phi - \Phi, R) ,
 \label{eq:momentum-vs-R}
 \\
 &&
 {\boldsymbol {\mathcal{A}}} (\phi - \Phi, R) =
 \frac{g \mu_B \lambda}{2 \hbar c} \, \boldsymbol{\mathcal E} (\phi,{\bf  R})  \times{\bf e}_{\phi}
 \nonumber \\ && \quad \quad =
 \frac{\lambda}{\lambda_0} \,
 \frac{r_0 \, (r_0 + R \cos (\phi - \Phi)}{r_0^2 + R^2 + 2 r_0 R \cos (\phi - \Phi)} \,
 {\bf e}_{z} .
 \label{eq:B-vector-vs-R}
\end{eqnarray}
For example, for neutrons, $g = -1.041 875 63 (25) \times 10^{-3}$ \cite{CODATA-18}, so $|\lambda_{0}^{\rm (n)}| = 6.544 \times 10^6$~esu/cm.  For alkali atoms, with $L = 0$ and $S = \frac{1}{2}$ in the ground state, $g \approx 2$, and $\lambda_{0}^{\rm (a)} = 3.409 \times 10^3$~esu/cm.
For a charged rod of radius $b = 1 ~ \mu$m and charge density $\rho = 2998 ~ \text{esu/cm}^3$, the linear charge density is $\lambda = 9.418 \times 10^{-5}$~esu/cm.   Therefore the ratios are $\lambda / \lambda_{0}^{\rm (a)} = 2.763 \times 10^{-8}$ and $\lambda / |\lambda_{0}^{\rm (n)}| = 1.439 \times 10^{-11}$, respectively; they are really small.

Section~\ref{subsec:properties-of-B} computes ${\boldsymbol {\mathcal{A}}} (\phi - \Phi, R)$ in Cartesian coordinates, and shows that ${\boldsymbol {\mathcal{A}}} (\phi - \Phi, R)$ depends on the vectors ${\bf r}$ and ${\bf R}$ (not on a single vector ${\bf r} - {\bf R}$).  Note that ${\boldsymbol {\mathcal{A}}} (\phi - \Phi, R)$ is directed along the $z$-axis, so $H ({\bf R})$ in Eq.~(\ref{eq:H-vs-R}) commutes with $\sigma_z$, but not with $\sigma_x$ and $\sigma_y$.  Therefore, the quantum states of the particle moving along the path $P$ can be parameterized by the spin projection on the $z$-axis, $\sigma$.  Furthermore, the vector field ${\boldsymbol {\mathcal{A}}} (\phi - \Phi, R)$ and the Hamiltonian $H ({\bf R})$ are invariant with respect to the transformations
\begin{equation}   \label{eq:B0symmetry-R}
\phi \to \phi' = \phi + \alpha ,
\quad
\Phi \to \Phi' = \Phi + \alpha ,
\end{equation}
where $\alpha$ is real.  The invariance of the Hamiltonian with respect to these transformations implies that the wave function of the particle is also invariant with respect to these transformations, i.e., they have the form $\psi_{\sigma} (\phi - \Phi, R)$. Furthermore, the energies depend solely on $R$, but not on $\Phi$.  The wave function $\psi_{\sigma} (\phi - \Phi, R)$ and energy $\epsilon$ are determined by the Schr\"odinger equation
\begin{equation}   \label{eq:Schrodinger-R}
H ({\bf R}) \, \psi_\sigma (\phi - \Phi, R) = \epsilon \, \psi_\sigma (\phi - \Phi, R) .
\end{equation}
Applying the unitary transformation with the unitary matrix $\mathcal{U} (\phi - \Phi, R)$,
\begin{equation}   \label{eq:U-vs-R}
\mathcal{U} (\phi - \Phi, R) = 
\exp \bigg[ - i \sigma_z \int\limits_{0}^{\phi - \Phi} \mathcal{A} (\phi', R) \, d\phi' \bigg] ,
\end{equation}
such that
\begin{eqnarray}
&&
H ({\bf R}) =
\mathcal{U} (\phi - \Phi, R) \, \tilde{H} \, \mathcal{U}^{-1} (\phi - \Phi, R) ,
\label{eq:H-transformation}
\\
&&
\psi_{\sigma} (\phi - \Phi, R) =
\mathcal{U} (\phi - \Phi, R) \, \tilde\psi_{\sigma} (\phi - \Phi) ,
\label{eq:psi-transformation}
\qquad
\end{eqnarray}
the transformed Hamiltonian takes the form,
\begin{eqnarray}   \label{eq:H-transformed}
\tilde{H} =
\frac{p^2}{2 M} ,
\end{eqnarray}
and the transformed wave function satisfies
\begin{equation}   \label{eq:Schrodinger-transformed-R}
\tilde{H} \, \tilde\psi_{\sigma} (\phi - \Phi) = \epsilon \, \tilde\psi_{\sigma} (\phi - \Phi) .
\end{equation}

The solution for the wave function in Eq.~(\ref{eq:Schrodinger-transformed-R}) is
\begin{equation}   \label{eq:wave-fun-perturb}
\tilde\psi_{m, \sigma} (\phi) =
\frac{e^{i k_{m, \sigma} \phi}}{\sqrt{2 \pi}} \, \chi_{\sigma} ,
\end{equation}
where $\chi_{\sigma}$ is the eigenvector of $\sigma_z$ with eigenvalue $\sigma$.  The azimuthal wave-number $k_{m, \sigma}$ is found from the boundary condition,
\begin{eqnarray}
\tilde\psi_{m, \sigma} (2 \pi) &=&
\mathcal{U}^{-1} (2 \pi, R) \, \tilde\psi_{m, \sigma} (0)
\nonumber \\ &=&
e^{i \sigma \, \varphi_{\rm AC}} \, \tilde\psi_{m, \sigma} (0) ,
\end{eqnarray}
where the AC phase, $\varphi_{\rm AC}$, does not depend on $R$.
The solution to this equation for $k_{m, \sigma}$ is given by $k_{m, \sigma} = m + \frac{\sigma}{2 \pi} \, \varphi_{\rm AC}$, where $m$ is an integer.
The corresponding energy is
\begin{equation}
\epsilon_{m, \sigma} = \frac{\hbar^2 k_{m, \sigma}^2}{2 M r_0^2} =
\frac{\hbar^2}{2 M r_0^2} \, \Big( m + \frac{\sigma}{2 \pi} \, \varphi_{\rm AC} \Big)^{2} .
\end{equation}

\subsection{Cartesian components of ${\boldsymbol {\mathcal{A}}} (\phi - \Phi, R)$ in Eq.~(\ref{eq:B-vector-vs-R})}  \label{subsec:properties-of-B}

The electric field ${\bf E} (\phi, {\bf R})$ in Eq.~(\ref{eq:EF-R}) can be rewritten using the Cartesian coordinates of ${\bf r} (\phi)$ and ${\bf R}$:
\begin{eqnarray}   \label{eq:EF-R-Cartesian}
{\bf E}(\phi, {\bf R}) &=&
\frac{2 \lambda}{[x (\phi) + X]^2 + [y (\phi) + Y]^2}
\nonumber \\ &\times&
\big[ (x (\phi) + X) \, \hat{\bf x} + (y (\phi) + Y) \, \hat{\bf y} \big] .
\end{eqnarray}
The equation demonstrates that the electric field depends solely on ${\bf r} (\phi) + {\bf R}$.
The unit vector ${\bf e}_{\phi}$ in the direction of the ring can be written as
\begin{equation}   \label{eq:e_phi-Cartesian}
{\bf e}_{\phi} = \frac{- y (\phi) \, \hat{\bf x} + x (\phi) \, \hat{\bf y}}{\sqrt{x^2 (\phi) + y^2 (\phi)}} ,
\end{equation}
and the vector ${\boldsymbol {\mathcal{A}}} (\phi - \Phi, R)$ in Eq.~(\ref{eq:B-vector-vs-R}) takes the form
\begin{eqnarray}
{\boldsymbol {\mathcal{A}}} (\phi - \Phi, R) =
\frac{\lambda}{\lambda_0} \,
\frac{x (\phi) [x (\phi) + X] + y (\phi) [y (\phi) + Y]}{[x (\phi) + X]^2 + [y (\phi) + Y]^2} \,
\hat{\bf z} ,
\nonumber \\
\label{eq:B-vector-vs-R-Cartesian}
\end{eqnarray}
where $\lambda_0$ is given in Eq.~(\ref{eq:lambda_0}).  This equation demonstrates that ${\boldsymbol {\mathcal{A}}} (\phi - \Phi, R)$ depends on ${\bf r}$ and ${\bf R}$, but not solely on ${\bf r} (\phi) + {\bf R}$.

\section{Connection between the AC phase and the Berry phase}  \label{sec:AC_phase_Berry_phase}

In this section we show that the AC phase is a Berry phase.  Note that Ref.~\cite{Reuter_91} has already mentioned that the AC phase is a special case of a Berry phase. The AC phase and the Berry phase are conceptually distinct in that the former involves an integral over the path in coordinate space, ${\bf r}$, while the latter is over a path in parameter space of the system, ${\bf R}$.  However, given that the AC phase is a function of ${\bf r} - {\bf R}$, the integral over ${\bf R}$ can be converted into an integral over ${\bf r}$.  Consequently, the AC phase is essentially a Berry phase, as will be explicitly demonstrated below.

\subsection{Berry connection and Berry phase}   \label{subsec:Berry-Berry}

The Berry connection for a particle with a magnetic moment propagating on the path $P$ shown in Fig.~\ref{Fig:AC-phase-Berry-phase} is given by (see Refs.~\cite{Band-Avishai-QuantumMechanics, He_91})
\begin{equation}   \label{eq:Berry-connection-def}
{\bf A}_{\sigma} (\Phi, R) =
i \int\limits_{0}^{2 \pi}
\psi_{\sigma}^{\dag} (\phi - \Phi, R) \,
\nabla_{\bf R}
\psi_{\sigma} (\phi - \Phi, R) \,
d \phi .
\end{equation}
Using Eq.~(\ref{eq:psi-transformation}), the Berry connection can be expressed as
\begin{eqnarray}   \label{eq:Berry-connection-tilde-psi}
{\bf A}_{\sigma} (\Phi, R) &=&
i \int\limits_{0}^{2 \pi}
\tilde\psi_{\sigma}^{\dag} (\phi - \Phi) \,
\nabla_{\bf R}
\tilde\psi_{\sigma} (\phi - \Phi) \,
d \phi
\nonumber \\ &+&
\sigma \int\limits_{0}^{2 \pi} d \phi \,
\tilde\psi_{\sigma}^{\dag} (\phi - \Phi) \,
\tilde\psi_{\sigma} (\phi - \Phi)
\nonumber \\ && \times
\nabla_{\bf R} \int\limits_{0}^{\phi - \Phi} d \phi' \, \mathcal{A} (\phi', R) ,
\end{eqnarray}
where $\mathcal{A} (\phi', R)$ is the $z$-component of $\boldsymbol{\mathcal{A}} (\phi', R)$ given in Eq.~(\ref{eq:B-vector-vs-R}), and $\nabla_{\bf R} = (\frac{\partial}{\partial R},  \frac{1}{R}\frac{\partial}{\partial \Phi})$ in polar coordinates.  The Berry connection can be expressed in terms of the polar components as follows:
\begin{equation}
{\bf A}_{\sigma} (\Phi, R) =
\big( A_{\sigma, R} (R), A_{\sigma, \Phi} (R) \big) ,
\end{equation}
where
\begin{eqnarray}
A_{\sigma, R} (R) &=&
\sigma \int\limits_{0}^{2 \pi}
\Big( 1 - \frac{\phi}{2 \pi} \Big) \,
\frac{\partial \mathcal{A} (\phi, R)}{\partial R} \, d \phi ,
\label{eq:Berry-connection-cylindrical-R}
\\
A_{\sigma, \Phi} (R) &=& - \frac{k_m}{R} - 
\frac{\sigma}{2 \pi R} \int\limits_{0}^{2 \pi} \mathcal{A} (\phi, R) \, d \phi .
\label{eq:Berry-connection-cylindrical-Phi}
\end{eqnarray}
Note that $A_{\sigma, R} (R)$ and $A_{\sigma, \Phi} (R)$ depend on $R$, but not on $\Phi$, because the variable $\phi - \Phi$ appears in the integrand of Eq.~(\ref{eq:Berry-connection-tilde-psi}) and therefore $\Phi$ can be transformed away.  Moreover, the integrals in Eqs.~(\ref{eq:Berry-connection-cylindrical-R}) and (\ref{eq:Berry-connection-cylindrical-Phi}) do not depend on $\sigma$; the only $\sigma$ dependence comes from the factor $\sigma$ multiplying the integrals.  Also note that the term $- k_m/R$ which appears in Eq.~(\ref{eq:Berry-connection-cylindrical-Phi}) is the wave number of the particle propagating on the ring; it is not related to the Berry phase calculated below.

The Berry phase is defined as \cite{Band-Avishai-QuantumMechanics}
\begin{eqnarray}   \label{eq:Berry-phase-def}
\gamma_{B, \sigma} (R) &=& - \oint\limits_{C} d {\bf R} \cdot {\bf A}_{\sigma} (R)
\nonumber \\ &=&
- \sigma \, \varphi_{\rm AC} (R) ,
\end{eqnarray}
where
\begin{eqnarray}   \label{eq:integral-B}
\varphi_{\rm AC} (R) &=&
\int\limits_{0}^{2 \pi} \mathcal{A} (\phi, R) \, d \phi
\nonumber \\ &=&
\left\{
  \begin{array}{ccc}
    \varphi_{\rm AC} & \text{for} & R < r_0 ,
    \\
    0 & \text{for} & R > r_0 ,
  \end{array}
\right.
\end{eqnarray}
and $\varphi_{\rm AC} = 2 \pi \, \frac{\lambda}{\lambda_0}$ is the AC phase.

\subsection{Berry curvature and Chern Number}  \label{subsec:Berry-Chern}

The Berry curvature is defined as \cite{Band-Avishai-QuantumMechanics}
\begin{equation}
\Omega_{\sigma} (R) = \nabla_{\bf R} \times {\bf A}_{\sigma} (\Phi, R) ,
\end{equation}
where
$$
\nabla_{\bf R} \times (f(R, \Phi), g(R, \Phi))  =
\frac{1}{R}
\Big\{
  \frac{\partial [ R g(R, \Phi)]}{\partial R} -
  \frac{\partial f(R, \Phi)}{\partial \Phi}
\Big\}
$$
in polar coordinates, and $(f, g)$ are polar components of ${\bf A}_{\sigma}$, hence
\begin{equation}   \label{eq:Berry-curvature-def}
\Omega_{\sigma} (R) =  \frac{1}{R} \, \frac{\partial [R \, {\mathcal A}_{\sigma, \Phi} (R)]}{\partial R} .
\end{equation}
The integral over $\phi$ on the right-hand-side of Eq.~(\ref{eq:Berry-connection-cylindrical-Phi}) is calculated in Eq.~(\ref{eq:integral-B}), and yields an expression for $\Omega_{\sigma} (R)$ having the following form:
\begin{equation}   \label{eq:Berry-curvature-res}
\Omega_{\sigma} (R) = - \frac{\sigma}{R} \, \frac{\lambda}{\lambda_0} \, \delta (R - r_0) .
\end{equation}

The Chern number $C$ for the AC effect is defined as \cite{Chern_46}
\begin{equation}   \label{eq:Chern-number-def}
C = \sum_{\sigma} C_{\sigma} ,
\end{equation}
where
\begin{eqnarray}
C_{\sigma} &=&
\frac{1}{2 \pi} \int\limits_{0}^{2 \pi} d\Phi \int\limits_{0}^{R(\Phi)} \Omega_{\sigma} (R) \, R \, dR ,
\label{eq:C_sigma}
\end{eqnarray}
where the equation $R = R(\Phi)$ defines the contour $C$ shown in Fig.~\ref{Fig:AC-phase-Berry-phase}, and the function $ R(\Phi)$ satisfies the conditions: periodicity, $R(\Phi + 2 \pi) = R(\Phi)$, and $R(\Phi) < r_0$.  Consequently, $C_{\sigma} = 0$  and the Chern number for the AC effect vanishes, $C = 0$.  However, if $R(\Phi) > r_0$, $C_\sigma \ne 0$ but Eqs.~(\ref{eq:Berry-connection-cylindrical-Phi}), (\ref{eq:Berry-curvature-def}) and (\ref{eq:C_sigma}) show that $C_{-1} = - C_{1}$, therefore the Chern number for the AC effect vanishes, $C = 0$, also for this case \cite{loverl0}.

\section{AC phase factor in a homogeneous electric field}  \label{sec:homogeneous-E-field}

\begin{figure}
\centering
  \includegraphics[width=0.8\linewidth,angle=0] {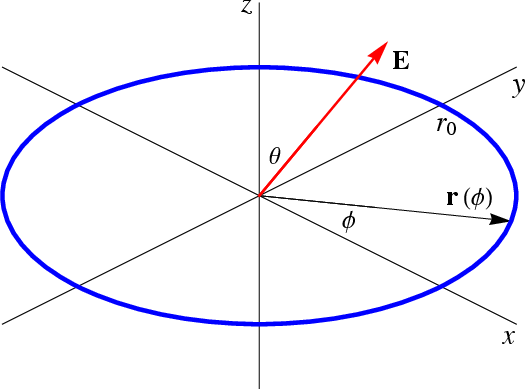}
\caption{\footnotesize The circular ring (blue) is in the $x$-$y$ plane and is defined by the vector ${\bf r}(\phi)$ in Eq.~(\ref{eq:r-phi}).  The radius of the ring is $r_0$, and the center of the ring is chosen as an origin of the coordinate system.  The ring is placed in a homogeneous electric field ${\bf E}$ which lies in the $x$-$z$ plane and makes an angle of $\theta$ with the $z$-axis, see Eq.~(\ref{eq:E-field}).}
\label{Fig:circle-Efield-homogeneous}
\end{figure}

Let us consider the motion of a neutral particle with spin-1/2 that moves along a ring of radius $r_0$ situated in the $x$-$y$ plane, and subject to a homogeneous external electric field
\begin{equation}   \label{eq:E-field}
 {\bf E} = E_0 (\sin \theta \, \hat{\bf x} + \cos \theta \, \, \hat{\bf z}) ,
\end{equation}
as shown in Fig.~\ref{Fig:circle-Efield-homogeneous}.
The ring is defined by the vector ${\bf r}(\phi)$ in Eq.~(\ref{eq:r-phi}).  The Pauli Hamiltonian is
\begin{eqnarray}   \label{eq:H_AC2}
H &=&
\frac{1}{2 M} \,
\Big( p + \frac{\hbar}{r_0} \, \boldsymbol\sigma \cdot \boldsymbol {\mathcal{A}} (\phi) \Big)^{2} ,
\end{eqnarray}
where
\begin{equation}   \label{eq:Avector-def}
  \boldsymbol{\mathcal A} (\phi) =
  \frac{r_0}{2 \lambda_0} \, {\bf E} \times {\bf e}_{\phi} = \mathcal{A}_{0}( -  \, \cos \theta \, {\bf e}_{r}
  + \sin \theta \, \cos \phi \, {\bf e}_{z}) ,
\end{equation}

$\lambda_0$ is given in Eq.~(\ref{eq:lambda_0}), and
\begin{equation}     \label{eq:A_0}
  \mathcal{A}_{0} = \frac{E_0 \, r_0}{2 \lambda_0} .
\end{equation}

The matrix $\boldsymbol\sigma \cdot \boldsymbol {\mathcal{A}} (\phi)$ can be written as
\begin{equation}   \label{eq:B-dot-sigma}
  \boldsymbol\sigma \cdot \boldsymbol {\mathcal A} (\phi) =
  \mathcal{A}_{0} \, \boldsymbol\sigma \cdot \boldsymbol \alpha (\phi) ,
\end{equation}
where
\begin{equation}     \label{eq:alpha-def}
  \boldsymbol\alpha (\phi) \equiv
  \frac{\boldsymbol{\mathcal A} (\phi)}{\mathcal{A}_{0}} =
  - \cos \theta \, {\bf e}_{r} + \sin \theta \, \cos \phi \, {\bf e}_{z} ,
\end{equation}
therefore,
\begin{equation}          \label{eq:sigma-dot-alpha}
  \boldsymbol\sigma \cdot \boldsymbol\alpha (\phi) =
  \left(
    \begin{array}{cc}
      \sin \theta \, \cos \phi & - cos \theta \, e^{- i \phi}
      \\
      - \cos \theta \, e^{i \phi} & - \sin \theta \, \cos \phi
    \end{array}
  \right) .
\end{equation}

The AC phase factor $\mathcal{U} (\phi)$ satisfies the equation
\begin{equation}   \label{eq:for-AC-phase-factor}
  \Big( p + \frac{\hbar}{r_0} \, {\mathcal A}_{0} \, \boldsymbol\sigma \cdot \boldsymbol\alpha (\phi) \Big) \,
  \mathcal{U} (\phi) = 0 ,
\end{equation}
and the boundary condition
\begin{equation}   \label{eq:U-boundary}
  \mathcal{U} (0) = \sigma_0 ,
\end{equation}
where $\sigma_0$ is the 2$\times$2 identity matrix.
The solution to Eq.~(\ref{eq:for-AC-phase-factor}) with the boundary condition in Eq.~(\ref{eq:U-boundary}) can be expressed as an ordered exponential,
\begin{eqnarray}   \label{eq:U-Pexp}
  \mathcal{U} (\phi) &=&
  {\mathcal P}\!\exp \!
  \bigg[
    - i \mathcal{A}_{0}
    \int\limits_{0}^{\phi} \boldsymbol\sigma \cdot \boldsymbol\alpha (\phi') \, d \phi'
  \bigg]
  \nonumber \\ &\equiv&
  \sum_{n = 0}^{\infty}
  \big( - i \mathcal{A}_{0} \big)^{n}
  \prod_{l = 1}^{n}
  \int\limits_{0}^{\phi_{l - 1}} \boldsymbol\sigma \cdot \boldsymbol\alpha (\phi_l) \, d \phi_l ,
\end{eqnarray}
where $\phi_0 = \phi$  (see the upper limit of integration, $\phi_{l-1}$, in the factor with $l = 1$).

In subsection \ref{subsec:E-weak}, the electric field will be assumed to be weak, and the AC phase will be expanded in powers of $\mathcal{A}_{0}$.  It will be shown that, to lowest order in $\mathcal{A}_{0}$, the AC phase is quadratic in $\mathcal{A}_{0}$.  In subsections \ref{subsec:AC-phase-Ex} and \ref{subsec:AC-phase-Ez} the Aharonov-Casher phase is determined for $\theta = \pi/2$ and $\theta = 0$ without assuming a weak field. Then, in subsection \ref{subsec:AC-phase-numerical}, the AC phase will be numerically calculated as a function of $\mathcal{A}_{0}$ without assuming a weak field.

The Aharonov-Casher (AC) phase factor $\mathcal{U}_{\rm AC} = \mathcal{U} (2 \pi)$ is given by
\begin{equation}   \label{eq:U-vs-Phi}
  \mathcal{U}_{\rm AC} =
  e^{- i \boldsymbol\Phi \cdot \boldsymbol\sigma} =
  \sum_{n = 1}^{\infty}
  \frac{( - i )^n}{n !} \, \big( \boldsymbol\Phi \cdot \boldsymbol\sigma \big)^{n} ,
\end{equation}
where $\boldsymbol\Phi$ is a real vector, and the AC phase is defined as $\varphi_{\rm AC} = | \boldsymbol\Phi |$. 

\subsection{Weak electric field}   \label{subsec:E-weak}
 
For a weak electric field, $\mathcal{A}_{0} \ll 1$ [see Eq.~(\ref{eq:A_0})], the magnitude of the vector $\boldsymbol\Phi$ is small and the vector can be expanded in $\mathcal{A}_{0}$ as
\begin{equation}   \label{eq:Phi-series}
  \boldsymbol\Phi =
  \boldsymbol\Phi_{1} \, \mathcal{A}_{0} +
  \boldsymbol\Phi_{2} \, \mathcal{A}_{0}^{2} +
  O (\mathcal{A}_{0}^{3}) ,
\end{equation}
where $\boldsymbol\Phi_{1}$ and $\boldsymbol\Phi_{2}$ are unknown real vectors which will be determined below.   Therefore the AC phase factor $\mathcal{U}_{\rm AC}$ in Eq.~(\ref{eq:U-vs-Phi}) can be expanded in $\mathcal{A}_{0}$ as
\begin{eqnarray}   \label{eq:U-vs-Phi-expansion}
  \mathcal{U}_{\rm AC} &=&
  \sigma_0 -
  i \, \boldsymbol\Phi_{1} \cdot \boldsymbol\sigma \, \mathcal{A}_{0} -
  \Big[
    i \, \boldsymbol\Phi_{2} \cdot \boldsymbol\sigma +
    \frac{1}{2} \, \big( \boldsymbol\Phi_{1} \cdot \boldsymbol\sigma \big)^{2}
  \Big]
  \mathcal{A}_{0}^{2}
  \nonumber \\ && +
  O (\mathcal{A}_{0}^{3}) .
\end{eqnarray}
On the other hand, the AC factor $\mathcal{U}_{\rm AC} = \mathcal{U} (2 \pi)$, see Eq.~(\ref{eq:U-Pexp}) with $\phi = 2 \pi$, is expanded in $\mathcal{A}_{0}$ as
\begin{eqnarray}   \label{eq:U_AC-Pexp-series}
  \mathcal{U}_{\rm AC} &=&
  \sigma_0 -
  i \mathcal{A}_{0}
  \int\limits_{0}^{2 \pi} \big( \boldsymbol\sigma \cdot \boldsymbol \alpha (\phi_1) \big) \, d \phi_1 -
  \mathcal{A}_{0}^{2}
  \int\limits_{0}^{2 \pi} \big( \boldsymbol\sigma \cdot \boldsymbol \alpha (\phi_1) \big) \, d \phi_1
  \nonumber \\ && \times
  \int\limits_{0}^{\phi_1} \big( \boldsymbol\sigma \cdot \boldsymbol \alpha (\phi_2) \big) \, d \phi_2 +
  O (\mathcal{A}_{0}^{3}) .
\end{eqnarray}
Comparing Eqs.~(\ref{eq:U-vs-Phi-expansion}) and (\ref{eq:U_AC-Pexp-series}), we find that
\begin{eqnarray}
  \boldsymbol\Phi_{1} \cdot \boldsymbol\sigma
  &=&
  \int\limits_{0}^{2 \pi} \big( \boldsymbol\sigma \cdot \boldsymbol \alpha (\phi_1) \big) \, d \phi_1 ,
  \label{eq:for-Phi_1}
  \\
  \boldsymbol\Phi_{2} \cdot \boldsymbol\sigma
  &=&
  - i 
  \int\limits_{0}^{2 \pi} \big( \boldsymbol\sigma \cdot \boldsymbol \alpha (\phi_1) \big) \, d \phi_1 \,
  \int\limits_{0}^{\phi_1} \big( \boldsymbol\sigma \cdot \boldsymbol \alpha (\phi_2) \big) \, d \phi_2
  \nonumber \\ && +
  \frac{i}{2} \, \big( \boldsymbol\Phi_{1} \cdot \boldsymbol\sigma \big)^{2} .
  \label{eq:for-Phi_2}
\end{eqnarray}
Therefore,
\begin{eqnarray}
  \boldsymbol\Phi_{1} &=&
  \int\limits_{0}^{2 \pi} \boldsymbol \alpha (\phi_1) \, d \phi_1 ,
  \label{eq:Phi_1-res}
  \\
  \boldsymbol\Phi_{2} &=&
  \int\limits_{0}^{2 \pi} d \phi_1
  \int\limits_{0}^{\phi_1} d \phi_2 \,
  \boldsymbol\alpha (\phi_1) \times \boldsymbol\alpha (\phi_2) .
  \label{eq:Phi_2-res}
\end{eqnarray}
The application of Eq.~(\ref{eq:alpha-def}) to these equations shows that:
\begin{eqnarray}
  \boldsymbol\Phi_{1} &=& {\bf 0} ,
  \label{eq:Phi_1-res}
  \\
  \boldsymbol\Phi_{2} &=& - 2 \pi \cos \theta \, \big( \sin \theta \, \hat{\bf x} + \cos \theta \, \hat{\bf z} \big) .
  \label{eq:Phi_2-res}
\end{eqnarray}
Substitution of Eqs.~(\ref{eq:Phi_1-res}) and (\ref{eq:Phi_2-res}) into Eq.~(\ref{eq:Phi-series}) yields
\begin{equation}   \label{eq:Phi-res}
  \boldsymbol\Phi =
  - 2 \pi \, {\mathcal A}_{0}^{2} \, \cos \theta \, \big( \sin \theta \, \hat{\bf x} + \cos \theta \, \hat{\bf z} \big)  +
  O (\mathcal{A}_{0}^{3}) .
\end{equation}
Therefore, the AC phase can be expressed as follows:
\begin{equation}   \label{eq:AC-phase-res}
  \varphi_{\rm AC} =
  2 \pi \, {\mathcal A}_{0}^{2} \, \cos \theta  +
  O (\mathcal{A}_{0}^{3}) .
\end{equation}

In the next three subsections the assumption of weak field, ${\mathcal A}_{0} \ll 1$ is {\it not} made, i.e., the calculations of $\varphi_{\rm AC}$ are carried out for arbitrary ${\mathcal A}_{0}$.

\subsection{Aharonov-Casher phase for $\theta = \pi/2$}   \label{subsec:AC-phase-Ex}

The vector $\boldsymbol\alpha (\phi)$ in Eq.~(\ref{eq:alpha-def}),
for $\theta = \pi/2$,  is
\begin{equation}   \label{eq:alpha-E_x}
  \boldsymbol\alpha (\phi) =
  \cos \phi \, {\bf e}_{z} ,
\end{equation}
and the matrix $\boldsymbol\sigma \cdot \boldsymbol\alpha (\phi)$ in Eq.~(\ref{eq:sigma-dot-alpha}) takes the form
\begin{equation}   \label{eq:sigma-dot-alpha-Ex}
  \boldsymbol\sigma \cdot \boldsymbol\alpha (\phi) =
  \left(
    \begin{array}{cc}
      \cos \phi & 0
      \\
      0 & - \cos \phi
    \end{array}
  \right) =
  \cos \phi \, \sigma_z .
\end{equation}
The commutator of the matrices $\boldsymbol\sigma \cdot \boldsymbol\alpha (\phi)$ and $\boldsymbol\sigma \cdot \boldsymbol\alpha (\phi')$ vanishes for any values of $\phi$ and $\phi'$, therefore the AC phase factor $\mathcal{U} (\phi)$ in Eq.~(\ref{eq:U-Pexp}) is Abelian and can be expressed as
\begin{equation}   \label{eq:U-exp-Ex}
  \mathcal{U} (\phi) =
  \exp \!
  \bigg[
    - i \mathcal{A}_{0} \, \sigma_z 
    \int\limits_{0}^{\phi} \cos \phi'  \, d \phi'
  \bigg] .
\end{equation}

Using Eq.~(\ref{eq:U-vs-Phi}), the vector $\boldsymbol\Phi$ is found to be
\begin{equation}   \label{eq:AC-phase-Abelian}
  \boldsymbol\Phi =
  \mathcal{A}_{0} \, {\bf e}_{z} \, \int\limits_{0}^{2 \pi} \cos \phi \, d \phi  = {\bf 0} .
\end{equation}
Therefore the AC phase (the absolute value of $\boldsymbol\Phi$) vanishes for the electric field in the $x$-$y$ plane.  Note that these results were obtained for arbitrary ${\mathcal A}_{0}$ (no perturbation approximation was employed).

\subsection{Aharonov-Casher phase for $\theta = 0$}   \label{subsec:AC-phase-Ez}

The case with $\theta = 0$ was considered explicitly in Ref.~\cite{Avishai_19}. In this special case, the non-Abelian AC phase factor can be expressed as a product of a spin-rotation matrix and an Abelian phase factor, as we shall now show.  The vector $\boldsymbol\alpha (\phi)$ in Eq.~(\ref{eq:alpha-def}) is
\begin{equation}   \label{eq:alpha-E_z}
  \boldsymbol\alpha (\phi) = - {\bf e}_{r} ,
\end{equation}
and the matrix $\boldsymbol\sigma \cdot \boldsymbol\alpha (\phi)$ in Eq.~(\ref{eq:sigma-dot-alpha}) takes the form
\begin{equation}   \label{eq:sigma-dot-alpha-Ez}
  \boldsymbol\sigma \cdot \boldsymbol\alpha (\phi) =
  \left(
    \begin{array}{cc}
      0 & - e^{- i \phi}
      \\
      - e^{i \phi} & 0
    \end{array}
  \right) .
\end{equation}
The commutator of the matrices $\boldsymbol\sigma \cdot \boldsymbol\alpha (\phi)$ and $\boldsymbol\sigma \cdot \boldsymbol\alpha (\phi')$ for $\phi \neq \phi'$ is non-vanishing,
\begin{equation}
  \big[
    \boldsymbol\sigma \cdot \boldsymbol\alpha (\phi) , \,
    \boldsymbol\sigma \cdot \boldsymbol\alpha (\phi')
  \big] =
  - 2 i \sin (\phi - \phi') \, \sigma_y ,
\end{equation}
therefore the AC phase factor $\mathcal{U} (\phi)$ in Eq.~(\ref{eq:U-Pexp}) is non-Abelian.
However, there is an SU(2) transformation that transforms $\mathcal{U} (\phi)$ to an Abelian form.
Applying the unitary matrix
\begin{equation}
  \mathcal{R} (\phi) = \exp \Big( \frac{i \phi \sigma_z}{2} \Big) ,
\end{equation}
the matrix $\boldsymbol\sigma \cdot \boldsymbol\alpha (\phi)$ is reduced to Abelian form:
\begin{equation}
  \mathcal{R} (\phi) \, \boldsymbol\sigma \cdot \boldsymbol\alpha (\phi) \, \mathcal{R}^{\dag} (\phi) = - \sigma_x .
\end{equation}
The AC phase factor $\mathcal{U} (\phi)$ in Eq.~(\ref{eq:U-Pexp}) can be expressed as
\begin{equation}   \label{eq:U-tilde}
  \mathcal{U} (\phi) = \mathfrak{R} (\phi) \, \tilde{\mathcal U} (\phi) ,
\end{equation}
and the following equation for $\tilde{\mathcal U} (\phi)$ is obtained
\begin{equation}  \label{eq:AC_phase_factor}
  \Big( p - \frac{\hbar \sigma_z}{2 r_0} - \frac{\hbar}{r_0} \, \mathcal{A}_{0} \, \sigma_x \Big) \,
  \tilde{\mathcal U} (\phi) = 0 .
\end{equation}
The solution of this equation is
\begin{eqnarray}   \label{eq:tilde-U-res}
  \tilde{\mathcal U} (\phi) &=&
  \exp \Big( i \, \frac{\sigma_z + 2 \mathcal{A}_{0} \sigma_x}{2} \, \phi \Big) =
  \cos \Big( \sqrt{\mathcal{A}_{0}^{2} + \frac{1}{4}} \, \phi \Big)
  \nonumber \\ && +
  \, i \, \frac{\sigma_z + 2 \mathcal{A}_{0} \sigma_x}{\sqrt{4 \mathcal{A}_{0}^{2} + 1}} \,
  \sin \Big( \sqrt{\mathcal{A}_{0}^{2} + \frac{1}{4}} \, \, \phi \Big) .
\end{eqnarray}
Then, having the SU(2) matrix $\mathcal{U}_{\rm AC} = \mathcal{U} (2 \pi)$ in hand, the AC phase is given by
\begin{equation}   \label{eq:AC-phase-vs-Tr}
  \varphi_{\rm AC} =
  \arccos \Big( \frac{1}{2} \, {\rm Tr} \big[ \mathcal{U}_{\rm AC} \big] \Big) = \pi \, \Big(  \sqrt{1 + 4 \mathcal{A}_{0}^{2}} - 1  \Big) ,
\end{equation}
see also Ref.~\cite{Avishai_19}.  The series expansion of $\varphi_{\rm AC}$ is roughly quadratic for small $\mathcal{A}_{0}$ (see also Eq.~(\ref{eq:AC-phase-res}) with $\theta = 0$, and the red curve in Fig.~\ref{Fig:AC-phase-E0-theta} below):
\begin{equation}   \label{eq:AC-phase-expansion_small}
\varphi_{\rm AC} =
2 \pi \,
\big[
  \mathcal{A}_{0}^{2} -
  \mathcal{A}_{0}^{4} +
  O(\mathcal{A}_{0}^{6})
\big] .
\end{equation}
For large $\mathcal{A}_{0}$, the AC phase is roughly linear with  $\mathcal{A}_{0}$:
\begin{equation}   \label{eq:AC-phase-expansion_large}
\varphi_{\rm AC} =
2 \pi \,
\Big[
  \mathcal{A}_{0} -
  \frac{1}{2} +
  \frac{1}{8 \mathcal{A}_{0}} +
  O(\mathcal{A}_{0}^{-3})
\Big] .
\end{equation}

\subsection{Numerical calculation of the Aharonov-Casher phase}   \label{subsec:AC-phase-numerical}

This section presents the numerical calculations of the SU(2) matrix $\mathcal{U} (\phi)$ using Eq.~(\ref{eq:for-AC-phase-factor}) with the boundary condition in Eq.~(\ref{eq:U-boundary}) for some values of the parameters $\mathcal{A}_{0}$ and $\theta$, and presents the results for the AC phase.  The matrix $\mathcal{U}_{\rm AC} = \mathcal{U} (2 \pi)$ can be expressed as [see Eq.~(\ref{eq:U-vs-Phi}) with $\boldsymbol\Phi = \varphi_{\rm AC} \, {\bf b}$]:
\begin{equation}   \label{eq:U-vs-phi_AC-b}
  \mathcal{U}_{\rm AC} =
  e^{- i \varphi_{\rm AC} \, {\bf b} \cdot \boldsymbol\sigma} =
  \cos \varphi_{\rm AC} \, \sigma_0 -
  i \sin \varphi_{\rm AC} \, {\bf b} \cdot \boldsymbol\sigma ,
\end{equation}
where $\cos \varphi_{\rm AC}$ and $\sin \varphi_{\rm AC} \, {\bf b}$ are given by
\begin{eqnarray}
  \cos \varphi_{\rm AC} &=&
  \frac{1}{2} \, {\rm Tr} \big[ \mathcal{U}_{\rm AC} \big] ,
  \label{eq:cos-varphi_AC-vs-Tr}
  \\
  \sin \varphi_{\rm AC} \, {\bf b} &=&
  \frac{1}{2} \, {\rm Tr} \big[ \mathcal{U}_{\rm AC} \, \boldsymbol\sigma \big] .
  \label{eq:sin-varphi_AC-b-vs-Tr}
\end{eqnarray}
Having the SU(2) matrix $\mathcal{U}_{\rm AC}$ in hand, the AC phase can be calculated using Eq.~(\ref{eq:AC-phase-vs-Tr}).

\begin{figure}
\centering
  \includegraphics[width=0.8\linewidth,angle=0] {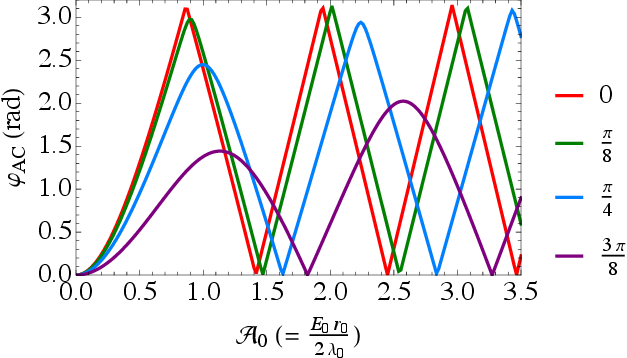}
\caption{\footnotesize The AC phase $\varphi_{\rm AC}$ numerically calculated as a function of $\mathcal{A}_{0}$ [see Eq.~(\ref{eq:A_0})] for various values of $\theta$.  For $\theta = \pi/2$, $\varphi_{\rm AC} = 0$.}
\label{Fig:AC-phase-E0-theta}
\end{figure}

Figure~\ref{Fig:AC-phase-E0-theta} shows the AC phase $\varphi_{\rm AC}$ versus $\mathcal{A}_{0}$ for various values of $\theta$.  For $\theta = 0$ (the red curve in Fig.~\ref{Fig:AC-phase-E0-theta}), the AC phase is given by the equation, $\varphi_{\rm AC} = \pi \, S(\mathcal{A}_{0})$, where $S(\mathcal{A}_{0}) = \sqrt{1 + 4 \mathcal{A}_{0}^{2}} - 1$, see Ref.~\cite{Avishai_19}.  The red curve in Fig.~\ref{Fig:AC-phase-E0-theta} is a piecewise function,
\begin{equation}   \label{eq:AC-phase-piecewise}
  \varphi_{\rm AC} =
  \left\{
    \begin{array}{ccc}
      \pi \, \big(  S(\mathcal{A}_{0}) - 2 n \big),
      &\text{for}&
      0 \leq S(\mathcal{A}_{0}) - 2 n < 1 ,
      \\
      \pi \big( 2 n - S(\mathcal{A}_{0})\big) ,
      &\text{for}&
      - 1 \leq S(\mathcal{A}_{0}) - 2 n < 0 ,
    \end{array}
  \right.
\end{equation}
where $n$ is integer.  For $\theta \neq 0$, $\varphi_{\rm AC}$ increases with $\mathcal{A}_{0}$, reaches its maximum smaller than $\pi$, and then decreases.  $\varphi_{\rm AC}$ depends on the angle $\theta$ and the ring radius $r_0$, therefore it is {\it a geometrical phase}, rather than a topological phase.   For $\theta = \pi/2$, the AC phase factor $\mathcal{U} (\phi)$ in Eq.~(\ref{eq:U-Pexp}) is Abelian, and the AC phase vanishes.

\section{Summary and Conclusion}  \label{sec:Summary}

It was demonstrated that the AC phase of a particle possessing a magnetic moment is not associated with the force exerted on the particle.  The AC Hamiltonian in Eq.~(\ref{eq:Foldy-Wouthuysen-Hamiltonian-general})  can be expressed in the form of the Hamiltonian in Eq.~(\ref{eq:PL}) for a static electric field.  The AC phase due to an electric field generated by a line of charge with a uniform linear charge density was calculated. The calculation was performed when the line of charge passed through the center of the ring and when it was off-center. In the former case, the line of charge can be at an angle $\theta$ from the normal to the plane of the ring, and in the latter case, the line of charge is normal to the plane of the ring. We demonstrated that the AC phase is a geometric Berry phase.  In general,, the AC phase factor is non-Abelian and the AC phase is not topological.  Our calculations have demonstrated that the AC phase for a particle having a magnetic moment and moving along a circular ring in the presence of a homogeneous electric field acquires an AC phase that is approximately quadratic with the electric field strength at small field strengths and is linear with the field strength at strong field strengths.



\begin{thebibliography}{99}
\bibitem{loverl0}  Recall that terms proportional to $\lambda^2/\lambda_0^2$ where neglected in the Hamiltonian in Eq.~(\ref{eq:H-vs-R}), which is used in the discussion of Berry connections, Berry curvatures and Chern numbers.  Nevertheless, the Chern number must vanish to all orders of $\lambda/\lambda_0$ because the full Hamiltonian for the AC effect is time-reversal invariant.
\end{thebibliography}
\end{document}